\documentstyle{article}
\font\tenrm=cmr10
\font\tenit=cmti10
\font\elevenbf=cmbx10 scaled\magstep 1
 1
 1

\def\Re{{\cal R \mskip-4mu \lower.1ex \hbox{\it e}\,}}
\def\Im{{\cal I \mskip-5mu \lower.1ex \hbox{\it m}\,}}

\def\etal{{\it et al.}}

\def\sub#1{_{\lower.25ex\hbox{$\scriptstyle#1$}}}
\def\sul#1{_{\kern-.1em#1}}
\def\sll#1{_{\kern-.2em#1}}
\def\sbl#1{_{\kern-.1em\lower.25ex\hbox{$\scriptstyle#1$}}}
\def\ssb#1{_{\lower.25ex\hbox{$\scriptscriptstyle#1$}}}
\def\sbb#1{_{\lower.4ex\hbox{$\scriptstyle#1$}}}

\def\gev{\,{\rm GeV}}

\def\to{\rightarrow}
\def\mh{\ifmmode m\sbl H \else $m\sbl H$\fi}
\def\mch{\ifmmode m_{H^\pm} \else $m_{H^\pm}$\fi}
\def\mt{\ifmmode m_t\else $m_t$\fi}
\def\mc{\ifmmode m_c\else $m_c$\fi}
\def\mz{\ifmmode M_Z\else $M_Z$\fi}
\def\mw{\ifmmode M_W\else $M_W$\fi}
\def\mws{\ifmmode M_W^2 \else $M_W^2$\fi}
\def\mhs{\ifmmode m_H^2 \else $m_H^2$\fi}
\def\mzs{\ifmmode M_Z^2 \else $M_Z^2$\fi}
\def\mts{\ifmmode m_t^2 \else $m_t^2$\fi}
\def\mcs{\ifmmode m_c^2 \else $m_c^2$\fi}
\def\mchs{\ifmmode m_{H^\pm}^2 \else $m_{H^\pm}^2$\fi}
\def\ztwo{\ifmmode Z_2\else $Z_2$\fi}
\def\zone{\ifmmode Z_1\else $Z_1$\fi}
\def\mtwo{\ifmmode M_2\else $M_2$\fi}
\def\mone{\ifmmode M_1\else $M_1$\fi}
\def\tb{\ifmmode \tan\beta \else $\tan\beta$\fi}
\def\xw{\ifmmode x\sub w\else $x\sub w$\fi}
\def\ch{\ifmmode H^\pm \else $H^\pm$\fi}
\def\lum{\ifmmode {\cal L}\else ${\cal L}$\fi}
\def\inpb{\ifmmode {\rm pb}^{-1}\else ${\rm pb}^{-1}$\fi}
\def\infb{\ifmmode {\rm fb}^{-1}\else ${\rm fb}^{-1}$\fi}
\def\epem{\ifmmode e^+e^-\else $e^+e^-$\fi}
\def\ppb{\ifmmode \bar pp\else $\bar pp$\fi}

\newskip\zatskip \zatskip=0pt plus0pt minus0pt
\def\matth{\mathsurround=0pt}
\def\lsim{\mathrel{\mathpalette\atversim<}}
\def\gsim{\mathrel{\mathpalette\atversim>}}
\def\atversim#1#2{\lower0.7ex\vbox{\baselineskip\zatskip\lineskip\zatskip
  \lineskiplimit 0pt\ialign{$\matth#1\hfil##\hfil$\crcr#2\crcr\sim\crcr}}}

\renewenvironment{thebibliography}[1]
 { \tenrm
   \begin{list}{\arabic{enumi}.}
    {\usecounter{enumi} \setlength{\parsep}{0pt}
     \setlength{\itemsep}{3pt} \settowidth{\labelwidth}{#1.}
     \sloppy
    }}{\end{list}}

\renewcommand{\thefootnote}{\fnsymbol{footnote}}

\begin{document} \begin{titlepage}
\rightline{\vbox{\halign{&#\hfil\cr
&ANL-HEP-CP-92-125\cr
&November 1992\cr}}}
\vspace{1in}
\begin{center}

{\Large\bf  CONSTRAINTS ON THE CHARGED HIGGS SECTOR FROM
               B PHYSICS }\footnote{Research supported by
the U.S. Department of
Energy, Division of High Energy Physics, Contract W-31-109-ENG-38.}

\bigskip

\normalsize
JOANNE L. HEWETT\\

\medskip
High Energy Physics Division\\
Argonne National Laboratory\\
Argonne, IL 60439\\

\end{center}
\bigskip
\bigskip

\begin{abstract}

We present the bounds that can be obtained on the charged Higgs sector
in two-Higgs-Doublet Models from measurements at LEP of the decay
$B\to D\tau\nu$, and from searches by CLEO for the inclusive decay
$b\rightarrow s\gamma$.

\end{abstract}

\vskip2.75in

\noindent{Presented at the {\it 1992 Meeting of the Division of Particles
and Fields},
Fermilab, Bataiva, IL, November 10-14, 1992. }

\renewcommand{\thefootnote}{\arabic{footnote}} \end{titlepage}

\begin{center}{{\elevenbf CONSTRAINTS ON THE CHARGED HIGGS SECTOR FROM
               B PHYSICS \\}
\vglue 0.6cm
{\tenrm J.L. HEWETT \\}
\baselineskip=13pt
{\tenit High Energy Physics Division, Argonne National Laboratory \\}
\baselineskip=12pt
{\tenit 9700 S.\ Cass Ave., Argonne, IL  60439  USA\\}
\vglue 0.2cm
{\tenrm ABSTRACT}}
\end{center}
\vglue 0.05cm
{\rightskip=3pc
 \leftskip=3pc
 \tenrm\baselineskip=12pt
 \noindent
We present the bounds that can be obtained on the charged Higgs sector
in two-Higgs-Doublet Models from measurements at LEP of the decay
$B\to D\tau\nu$, and from searches by CLEO for the inclusive decay
$b\rightarrow s\gamma$.
\vglue 0.6cm}
\tenrm
Many extensions of the Standard Model (SM) predict the existence of an
enlarged Higgs sector beyond the minimal one-doublet version\cite{hhg}.
The simplest extensions are models with two-Higgs-Doublets (2HDM), which
predict a physical spectrum of three neutral Higgs scalars, two of which
are CP-even ($h^0, H^0$) while one
is CP-odd ($A^0$), and two charged Higgs scalars ($H^\pm$).
We consider two distinct 2HDM which naturally avoid
tree-level flavor changing neutral currents.
In Model\ I, one doublet ($\phi_2$) provides masses for all fermions and the
other doublet ($\phi_1$) decouples from the fermion sector.
In a second model (Model\ II), $\phi_2$ gives mass to the up-type quarks, while
the down-type quarks and charged leptons receive their masses from $\phi_1$.
Supersymmetry and many axion theories predict couplings of the type present
in Model\ II.  Each doublet obtains a vacuum expectation value (vev) $v_i$,
subject only to the constraint that
$v_1^2 +v^2_2=v^2$, where $v$ is the usual vev present in the SM.
In a general 2HDM, the charged Higgs mass $m_{H^\pm}$
and the ratio of vevs, $v_2/v_1\equiv\tb$, are
{\it a priori} free parameters, as are the masses of all the neutral Higgs
fields.  However, in supersymmetric models, mass relationships exist between
the various Higgs scalars.  At tree-level, in such models,
only two parameters are required
to fix the masses and couplings of the entire scalar sector, but once radiative
corrections are included\cite{radcorr}, the values of the top-quark and squark
masses also need to be specified.

The strongest direct search limits on charged Higgs bosons are from
LEP\cite{lep}, with $\mch\gsim\mz/2$.  These charged
scalars may also reveal themselves through contributions to a variety of
low-energy processes.
Previous analyses\cite{bhp} of the \ch\ contributions to
processes such as $B^0-\bar B^0$, $D^0-\bar D^0$, and $K^0-\bar K^0$ mixing
have found the approximate bound $\tb\gsim\mt/600\gev$.
\ch\ bosons may also mediate the tree-level decay\cite{bdtn} $B\to D\tau\nu$.
In Model\ I, enhancements over the SM rate for this process only occur for
values of \tb\ which violate the above bound from meson anti-meson mixing.
However, enhancements are found in Model\ II for large values of \tb.
In Fig.\ 1 we show the ratio of $B(B\to D\tau\nu)$ in Model\ II to that of the
SM as a function of \tb, for various values of \mch.  The solid horizontal
line represents the $90\%$ C.L. upper bound on this ratio as obtained by
ALEPH\cite{lep}.  We see that for some range of the parameters, the value
of this ratio exceeds the experimental bound, and that very large values of
\tb\ are thus excluded.  For example, $\tb\lsim 60$ for $\mch=100\gev$.

Next we examine the radiative b-quark decay $b\to s\gamma$.   A $90\%$
C.L. upper bound on the branching fraction for this mode,
$B(b\rightarrow s\gamma)
< 8.4\times 10^{-4}$,  has been obtained by the CLEO Collaboration\cite{cleo}
via an examination of the inclusive photon spectrum in
$B$-meson decays.  Recent detector refinements coupled with increasing
integrated luminosity leads us to anticipate that either the current limit will
be strengthened, or the decay may actually be observed in the near future.
The transition $b\rightarrow s\gamma$ proceeds  through electromagnetic penguin
diagrams, which involve the top-quark, together with a $H^\pm$ or SM $W^\pm$
boson in the loop.
At the $W$ scale the coefficients of the operators which mediate this
transition
take the generic form\cite{bhp,bsg}
\begin{equation}
c_i(M_W)=A_W(m_t^2/M_W^2)+\lambda A^1_H(m_t^2/m_{H^\pm}^2)+
{1\over\tan^2\beta}A^2_H(m_t^2/m_{H^\pm}^2)  \,,
\end{equation}
where $\lambda = -1/\tan^2\beta,\ +1$\ in Models I and II, respectively, $A_W$
corresponds to the SM amplitude, and $A_H^{1,2}$ represent the $H^\pm$
contributions; their analytic form is given for each contributing operator in
Ref.\ 4,7.  We employ the explicit form of the QCD corrections  of Grinstein
{\it et al.}\cite{bsg}, and use the 3-loop expression for
$\alpha_s$ fitting the value of the QCD scale $\Lambda$ to obtain consistency
with measurements\cite{lep} of $\alpha_s(M_Z^2)$ at LEP.

Enhancements over the SM rate occur in Model\ I only for small
values of $\tan\beta$.
In Model\ II, large enhancements also appear
for small values of $\tan\beta$, but
more importantly, the branching fraction is found to {\it always} be larger
than that of the SM.
For certain ranges of the model parameters, the resulting value
of $B(b\rightarrow s\gamma)$ exceeds the CLEO bound,
and consistency with this limit
thus excludes part of the $m_{H^\pm} - \tan\beta$ plane for a
fixed value of $m_t$.  This
is shown in Fig.\ 2 for both models, where the excluded region lies to the left
and beneath the curves.
In Fig.\ 3a we present
the branching fraction for Model\ II in the limit of large $\tan\beta$
as a function of $m_t$\ with $m_{H^\pm}=m_t-m_b$.
If the actual value (or future upper bound) for
the branching fraction were to lie below the solid curve, then the decay
$t\rightarrow bH^\pm$ would be kinematically forbidden for
a particular value of $m_t$.  Here, we made use of the
large $\tan\beta$ limit, since it minimizes the $H^\pm$ contributions to
$b\rightarrow s\gamma$.

In the supersymmetric case, the bounds shown in Fig.\ 2b are more
conventionally displayed as an allowed region in the
$\tan\beta - m_A$\ plane, where
$m_A$ is the mass of the CP-odd field.  This is displayed in
Fig.\ 3b for various values of $m_t$, where the radiative corrections to the
SUSY mass relations have been employed assuming $M_{SUSY}=1\,{\rm TeV}$.
For $m_t=150\,{\rm GeV}$, the excluded region is comparable to what can be
explored by LEP I and II.
We note that in supersymmetric theories, other super-particles can also
contribute to the one-loop decay $b\rightarrow s\gamma$, and generally lead to
a further enhancement in the rate\cite{bert}.

In conclusion, we have shown that the decay $b\rightarrow s\gamma$ is by far
the most
restrictive process in constraining the parameters of the charged Higgs
sector in 2HDM, yielding bounds which are stronger than those from other
low-energy processes and from direct collider searches.
\\

{\elevenbf\noindent  References \hfil}
\vglue 0.2cm

\newpage

%
\noindent
{\tenrm {Fig. 1.  The ratio of branching fractions for $B\to D\tau\nu$ in
Model II to that of the SM.  The solid horizontal line represents the
ALEPH upper bound.  From left to right the solid (dashed-dot, dashed,
dotted, solid) curve represents $\mch=50\ (100, 150, 200, 250)\gev$.}
\vglue 0.2cm
\noindent
{\tenrm {Fig. 2.  The excluded regions in the $m_{H^\pm}-\tan\beta$ plane
for various
values of $m_t$, resulting from the present CLEO bound in (a) Model\ I and
(b) Model\ II.  In each case, from top to bottom, the solid
(dashed dot, solid, dotted, and dashed)  curve corresponds
to $m_t=210 (180, 150, 120,\ {\rm and}\ 90)\,{\rm GeV}$.
The excluded region lies to the left and below each curve.}}
\vglue 0.2cm
\noindent
{\tenrm {Fig. 3.  (a) $B(b\rightarrow s\gamma)$ as a function of $m_t$, with
$m_{H^\pm}=m_t-m_b$ in the large $\tan\beta$ limit in Model\ II.
(b)  The excluded region from the present CLEO limit in the
$\tan\beta- m_A$ plane for various values of $m_t$\ as indicated.}}

\end{document}